\documentclass{article}

\usepackage[final,nonatbib]{ml4eng_2020}
\usepackage[utf8]{inputenc} % allow utf-8 input
\usepackage[T1]{fontenc}    % use 8-bit T1 fonts
\usepackage[hidelinks]{hyperref}       % hyperlinks
\usepackage{url}            % simple URL typesetting
\usepackage{booktabs}       % professional-quality tables
\usepackage{amsfonts}       % blackboard math symbols
\usepackage{nicefrac}       % compact symbols for 1/2, etc.
\usepackage{microtype}      % microtypography
\usepackage{graphicx}
\usepackage{bm}
\usepackage{hyperref}
\usepackage{listings}

\usepackage{todonotes}
\usepackage{color}

\newcommand{\rs}[1]{\textcolor{black}{#1}}

\title{Multilevel Delayed Acceptance MCMC with an Adaptive Error Model in \texttt{PyMC3}}

\author{%
  Mikkel B.~Lykkegaard \\
  Centre for Water Systems and \\
  Institute for Data Science and Artificial Intelligence\\
  University of Exeter \\
  EX4 4QF, United Kingdom \\
  \texttt{m.lykkegaard@exeter.ac.uk} \\
  \And
  Grigorios ~Mingas \\
  The Alan Turing Institute  \\
  NW1 2DB, United Kingdom\\
  \texttt{gmingas@turing.ac.uk} \\
  \And
  Robert Scheichl \\
  Institute for Applied Mathematics and \\
  Interdisciplinary Center for Scientific Computing \\ 
  Ruprecht-Karls-Universität Heidelberg \\
  69120 Heidelberg, Germany \\
  \texttt{r.scheichl@uni-heidelberg.de} \\
  \And
  Colin Fox \\
  Department of Physics \\
  University of Otago \\
  Dunedin 9016, New Zealand \\
  \texttt{colin.fox@otago.ac.nz} \\
  \And
  Tim J.~Dodwell \\
  Institute for Data Science and Artificial Intelligence \\
  University of Exeter \\
  EX4 4QF, United Kingdom \\
  \texttt{t.dodwell@exeter.ac.uk} \\
}

\begin{document}

\maketitle
\begin{abstract}
  Uncertainty Quantification through Markov Chain Monte Carlo (MCMC) can be prohibitively expensive for target probability densities with expensive likelihood functions, for instance when the evaluation it involves solving a Partial Differential Equation (PDE), as is the case in a wide range of engineering applications. Multilevel Delayed Acceptance (MLDA) with an Adaptive Error Model (AEM) is a novel approach, which alleviates this problem by exploiting a hierarchy of models, with increasing complexity and cost, and correcting the inexpensive models on-the-fly. The method has been integrated within the open-source probabilistic programming package \texttt{PyMC3} and is available in the latest development version. In this paper, the algorithm is presented along with an illustrative example.
\end{abstract}

\section{Introduction}

Sampling from an unnormalised posterior distribution $\pi(\cdot)$ using Markov Chain Monte Carlo (MCMC) methods is a central task in computational statistics. This can be a particularly challenging problem when the evaluation of $\pi(\cdot)$ is computationally expensive and the parameter space $\theta$ and data ${\bf d}$ defining $\pi(\cdot)$ are high-dimensional. The sequential (highly) correlated nature of a Markov chain and the slow converge rates of Monte Carlo sampling, means that many MCMC samples are often required to obtain a sufficient representation of a posterior distribution $\pi(\cdot)$. Examples of such problems frequently occur in Bayesian inverse problems, image reconstruction and probabilistic machine learning, where simulations of the measurements (required to calculate a likelihood) depend on the evaluation of complex mathematical models (e.g. a system of partial differential equations) or the evaluation of prohibitively large data sets.

In this paper a MCMC approach capable of accelerating existing sampling methods is proposed, where a hierarchy (or sequence) $\pi_0(\cdot), \ldots, \pi_{L-1}(\cdot)$ of computationally cheaper approximations to the `full' posterior density $\pi(\cdot) \equiv \pi_L(\cdot)$ are available. As with the original delayed acceptance algorithm, proposed by Christen and Fox \cite{christen_markov_2005}, the idea is to generate MCMC proposals for the next step in the chain from runs of MCMC subchains targeting the computationally cheaper, approximate densities. The original DA method proposed the approach for just two levels. In this paper,  the approach is extended to recursively apply delayed acceptance across a complete hierarchy of model approximations, a method termed {\em multilevel delayed acceptance} (MLDA). There are close connections to and similarities with multilevel variance reduction techniques, first proposed by Giles \cite{giles_multilevel_2008}, widely studied for forward uncertainty propagation problems and importantly extended to Multilevel Markov Chain Monte Carlo approach by Hoang et al. \cite{hoang_mlmcmc_2013} and Dodwell {\em et al.} \cite{dodwell_hierarchical_2015}, and further to a Multi-Index setting by Jasra {\em et al.} \cite{jasra_multi-index_2017}. 
As in other multilevel approaches, the subchains in MLDA can be exploited for variance reduction, but this is beyond the scope of this paper.

The increase in use of Bayesian probabilistic tools has naturally coincided with the development of user-friendly computational packages, allowing users to focus on model development and testing, rather than algorithm development of sampling methods and post-processing diagnostics. Various high quality packages are available. Examples include: \texttt{MUQ}, \texttt{STAN} and \texttt{Pyro}.\footnote{\texttt{MUQ}: \href{http://muq.mit.edu}{http://muq.mit.edu}, \texttt{STAN}: \href{https://mc-stan.org}{https://mc-stan.org}, \texttt{Pyro}: \href{https://pyro.ai}{https://pyro.ai}} A guiding principle of our work and of this contribution was to ensure that the MLDA implementation is easily accessible, well supported and gives flexibility to users to define complex models in a friendly language. To achieve this we embed our sampler into the widely used open-source probabilistic programming package \texttt{PyMC3} \cite{salvatier_2016}. The method and implementation have been accepted in the development version, and will be made available with the next full release (version 3.9.4).

\section{Adaptive Multilevel Delayed Acceptance (MLDA)}

\subsection{Preliminaries: Metropolis-Hastings MCMC Algorithms}

Here, a typical Bayesian inverse problem is considered. Given are (limited) observations $d \in \mathbb R^M$ of a system and a mathematical model $\mathcal F(\theta): \mathbb R^R \mapsto \mathbb R^M$, which maps from a set of model parameters $\theta \in \mathbb R^R$ to the space of model predictions of the data. The connection between model and data is then, in the simplest case, described by the additive model
\begin{equation}
d = \mathcal F(\theta) + \epsilon
\end{equation}
(but it can also be more general).
Here, $\epsilon$ is a random variable, which can depend on $\theta$ and captures the uncertainty of the model's reproduction of the data. It might include measurement uncertainty of the recorded data, uncertainty due to model mis-specification and/or uncertainties due to sing in practice a numerical approximation of the mathematical model. The distribution of the random variable $\epsilon$ defines the {\em likelihood}, i.e. the probability distribution $\mathcal{L}(d|{\theta})$. For simplicity it is assumed to be Gaussian, i.e. \rs{$\epsilon \sim \mathcal{N}(\mu_\epsilon,\Sigma_\epsilon)$ and} $\mathcal{L}(d|\theta) \sim \mathcal{N}(d - \mathcal{F}(\theta) - \mu_\epsilon, \Sigma_\epsilon)$, but it does not have to be.

Given {\em prior} information $\pi(\theta)$ on the distribution of the model parameters ${\bf \theta}$, the aim is to condition this distribution on the observations, i.e. to obtain samples from the {\em posterior} distribution $\pi({\bf \theta}|{\bf d})$. Through Bayes' theorem, it follows that
\begin{equation}
\pi(\theta|d) = \frac{\mathcal{L}(d|\theta)\pi(\theta)}{\pi(d)} \propto \mathcal{L}(d|\theta)\pi(\theta).
\end{equation}
Since the normalising constant $\pi(d)$ (the {\em evidence}) is not typically known, the conditional distribution $\pi(\theta|d)$ is generally intractable and exact sampling is not possible. There are various computational strategies for generating samples from $\pi(\theta|d)$. This paper focuses on the Metropolis-Hastings MCMC algorithm, described in Algorithm 1.
It creates a Markov chain $\{\theta^{j}\}_{j \in \mathbb N}$ of correlated parameter states $\theta^{j}$ that (in the limit) target the exact posterior distribution $\pi(\theta|d)$ (cf.~e.g.~\cite{roberts2004}). The efficiency of the algorithm is determined by the \rs{choice} of the proposal distribution $q(\cdot | \cdot)$.

\begin{figure}[t]
    \fbox{\parbox{0.98\textwidth}{
            \textbf{Algorithm 1 (Metropolis-Hastings MCMC):} \ 
            %
            %\smallskip
            \rs{Choose $\theta^{0}$. Then, for $j=0,\ldots,J-1$:}\smallskip
            \begin{enumerate}
                \item Given $\theta^{j}$, generate a proposal $\theta'$ from a given proposal distribution $q(\theta'|\theta^{j})$,
                
                \item Accept proposal $\theta'$ as the next sample with probability
                 $$
                 \alpha(\theta'|\theta^j) = \min \left \{1, \frac{\mathcal{L}(d|\theta') \, \pi(\theta') \, q(\theta^{j}|\theta')}{\mathcal{L}(d|\theta^{j})\pi(\theta^{j})q(\theta'|\theta^{j})} \right\}\,,
                 $$
				i.e. set $\theta^{j+1} = \theta'$ with probability $\alpha$, and $\theta^{j+1} = \theta^{j}$ with probability $1 - \alpha$.             
            \end{enumerate}
    }}
\end{figure}

Whilst MCMC methods are the gold-standard for sampling from complex posterior distributions, for many types of models and data they come with significant practical challenges. Firstly, each cycle of Alg.~1 requires the evaluation of the model $\mathcal{F}(\theta')$ which may be computationally very expensive. Secondly, the samples generated in the chain are correlated, and therefore many cycles of Alg.~1 are often required to produce a sufficient number of ''independent'' (or \emph{effective}) samples from $\pi(\theta|{\bf d})$. The ideal proposal distribution generates cheap candidate proposals $\theta'$, which have a high probability of being accepted, and are independent of the previous sample $\theta^{j}$. 

\rs{In this paper, efficient, Metropolis-style proposal strategies are developed that} exploit a hierarchy of approximations $\mathcal{F}_\ell({\bf \theta})$, for $\ell = 0, \ldots, L-1$, to the full model $\mathcal{F}_L := \mathcal F$, which are assumed to be ordered according to increasing accuracy and computational cost.

\subsection{Multilevel Delayed Acceptance} 
Delayed Acceptance (DA) is an approach first introduced by Christen and Fox \cite{christen_markov_2005}, \rs{exploiting a simple, but highly effective idea}. The original DA approach is a two-level method that assumes a computationally cheaper approximation $\mathcal F^*$ for the forward map $\mathcal F$ is  available. The idea is that \rs{for any chosen proposal $\theta^\prime$, a standard Metropolis accept/reject step (as given in Alg.~1) is performed with the approximate forward map ${\mathcal F}^*(\theta^\prime)$ before the expensive forward model $\mathcal F(\theta^\prime)$ is evaluated}. Only if accepted, a second accept/reject step with the original forward map $\mathcal F (\theta^\prime)$ and with acceptance probability 
$\alpha = \min  \left\{ 1, \frac{\mathcal{L}(d|{\mathbf \theta}')  \mathcal{L}^*(d|{\mathbf \theta}^{j})}{\mathcal{L}(d|{\mathbf \theta}^{j}) \mathcal{L}^*(d|{\mathbf \theta}')} \right\}$
is carried out. \rs{Here,} $\mathcal{L}^*(d|\cdot)$ denotes the posterior distribution with the likelihood defined by $\mathcal{F}^*$. The validity of this approach as a proposal method, yielding a convergent MCMC algorithm, is provided in  \cite{christen_markov_2005}.

The basic DA approach can be extended in two ways. First, instead of doing a single check for the proposal that comes from the fine level, a subchain of length $J$ can be ran on the coarse level \cite{liu_2004,lykkegaard_accelerating_2020}. This does not affect the theory, but has the advantage of decorrelating samples passed back as proposals to the fine level. Second, and this is the main, novel algorithmic contribution, DA is extended to a general multilevel setting, exploiting links to the Multilevel Markov Chain Monte Carlo (MLMCMC) Method proposed by Dodwell {\em et al.} \cite{dodwell_hierarchical_2015}. 

\begin{figure}[ht]
    \centering
    \includegraphics[height=2.8cm]{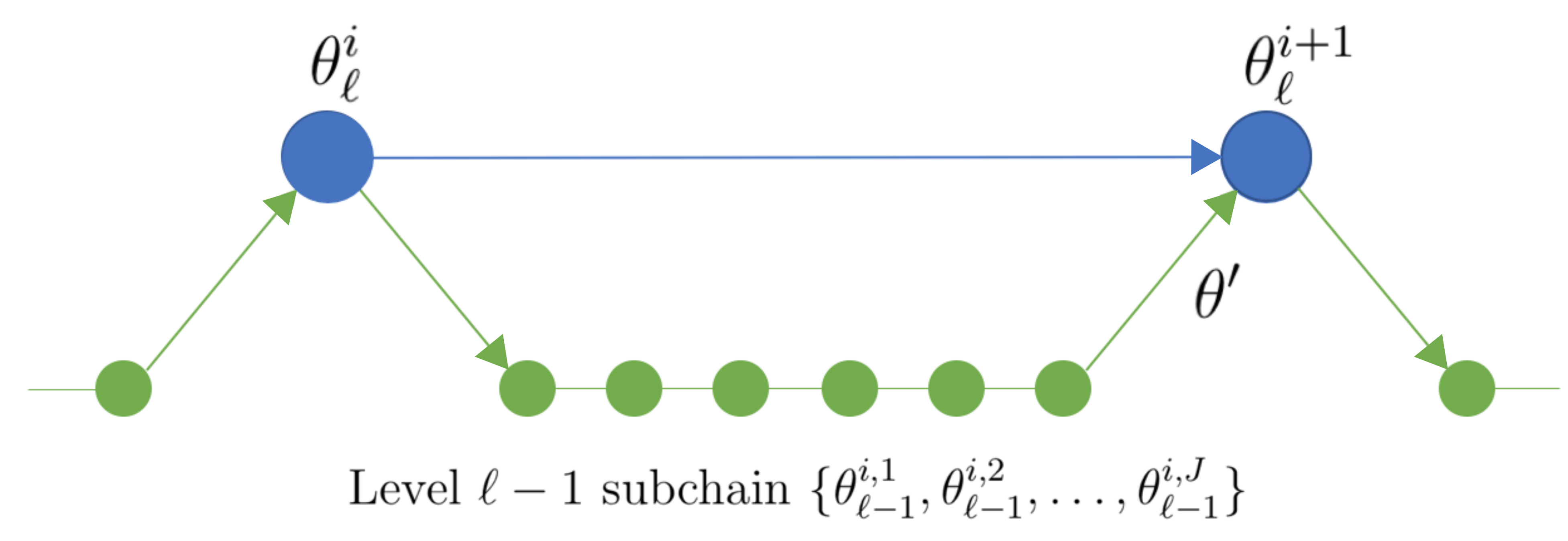}
    \includegraphics[height=2.8cm]{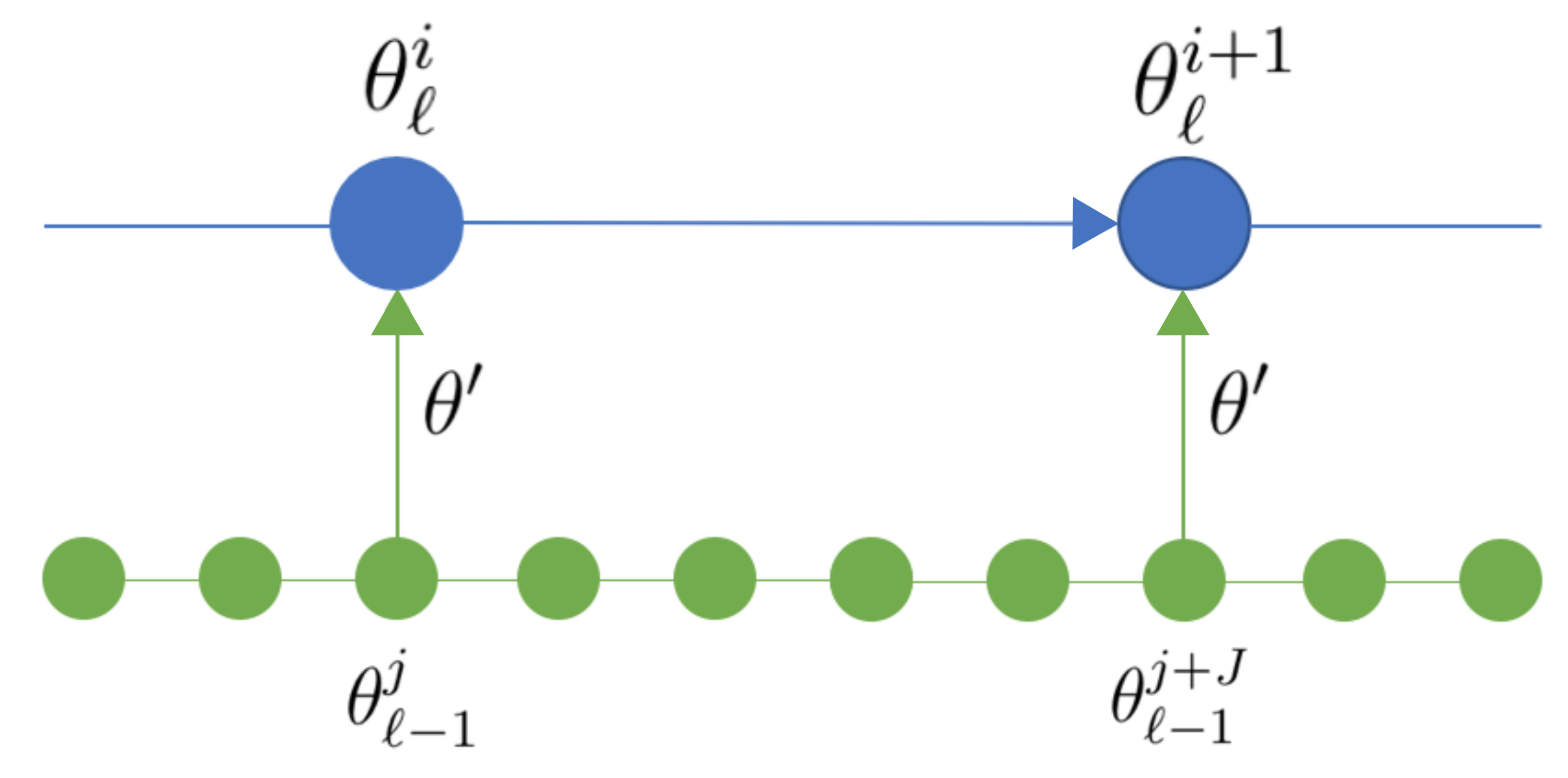}
    \caption{\rs{Schematic for generating a proposal $\theta^\prime$ on level $\ell$ in MLDA (left) and in MLMCMC (right).}} 
    %The key difference is that in MLMCMC, proposals for the coarse chain on level $\ell-1$ are generated independently of the state on level $\ell$.}
    \label{fig:mlda_vs_mlmcmc}
\end{figure}

The subtle differences between \rs{the approaches are apparent when comparing the schematics of the two multilevel proposal processes} shown in Fig.~\ref{fig:mlda_vs_mlmcmc}.
\rs{Algorithmically, Multilevel Delayed Acceptance (MLDA)} can be seen as a recursion of Delayed Acceptance over multiple levels $\ell = \{0, 1, \dots, L\}$. Crucially, if $\theta_\ell^{i}$ is the current state at level $\ell$, and a proposal $\theta'$ from the coarse subchain on level $\ell-1$ is rejected at level $\ell$, the coarse subchain to generate the subsequent proposal for level $\ell$ is again initiated from $\theta_\ell^{i}$. For MLMCMC, even if the coarse proposal is rejected, the coarse chain continues independently of the fine chain and does not revert to the state $\theta_\ell^{i}$ \rs{(see Fig.~\ref{fig:mlda_vs_mlmcmc}, right)}. As a result, coarse and fine 
chains will detach, and only align once a coarse proposal is accepted at the fine level.

\rs{The new MLDA algorithm with subchain length $J_\ell \in \mathbb N$ on level $0 \le \ell < L$ is described in Algorithm~2.}

\begin{figure}[th]
     \fbox{\parbox{0.98\textwidth}{
            \textbf{Algorithm 2 (Multilevel Delayed Acceptance MCMC):}
            
            \smallskip
            \rs{Choose $\theta^{0}$ and set the states of all subchains $\theta_0^0 = \ldots = \theta_{L-1}^0 = \theta^0$. Then, for $j=0,\ldots,J-1$:
            \begin{enumerate}
                \item Given $\theta^{j}$ and $\theta_\ell^{j_\ell}$ %(the current states of all chains) 
                such that $j_\ell < J_\ell$ for all $1 \le \ell < L$, generate a subchain of length $J_0$ with Alg.~1 on level $0$, starting from $\theta_0^0 = \theta_1^{j_1}$ and using the transition kernel $q({\theta}'_0 | {\theta}_0^{j_1})$. 
                \item Let $\ell = 1$ and $\theta'_1 = \theta_{0}^{J_{0}}$.
                \item %If $\ell = L$ go to Step 3; otherwise set $\theta'_\ell = \theta_{\ell-1}^{J_{\ell-1}}$ and 
                If $\ell = L$ go to Step 7. Otherwise compute the delayed acceptance probability on level $\ell$, i.e.,\vspace{-1ex}
                \[
                %\textstyle 
                \alpha_\ell = \min  \left\{ 1, \frac{\mathcal{L}_{\ell}(d | \theta'_\ell)\; \mathcal{L}_{\ell-1}(d | \theta_\ell^{j_\ell}) }{\mathcal{L}_{\ell}(d | \theta_\ell^{j_\ell}) \mathcal{L}_{\ell - 1}(d | \theta'_\ell)\;\,} \right\}.
                \]
                \item Set $\theta_\ell^{j_\ell+1} = \theta'_\ell$ with probability $\alpha_\ell$ and $\theta_\ell^{j_\ell+1} = \theta_\ell^{j_\ell}$ otherwise. Increment $j_\ell \to j_{\ell}+1$.
                \item If $j_\ell = J_\ell$ set $\theta'_{\ell+1} = \theta_{\ell}^{J_{\ell}}$, increment $\ell \to \ell+1$ and return to Step 3.
                \item Otherwise set $j_k = 0$ and $\theta^0_k = \theta^{j_\ell}_\ell$, for all $0 \le k < \ell$, and return to Step 1.
                \item Compute the delayed acceptance probability on level $L$, i.e.,
                \[
                %\textstyle 
                \alpha_L = \min  \left\{ 1, \frac{\mathcal{L}_{\ell}(d | \theta_L') \mathcal{L}_{\ell-1}(d | \theta^{j})\, }{\mathcal{L}_{\ell}(d | \theta^{j})\, \mathcal{L}_{\ell - 1}(d | \theta_L')} \right\}.
                \]
                Set $\theta^{j+1} = \theta_L'$ with probability $\alpha_L$ and $\theta^{j+1} = \theta^{j}$ otherwise. Increment $j \to j+1$. 
                \item Set $j_\ell = 0$ and $\theta^0_\ell = \theta^{j}$, for all $0 \le \ell < L$, and return to Step 1.
            \end{enumerate}
    }}}
\end{figure}

%\todo[inline]{Rob: Here is my version of the MLDA algorithm, which I think is correct, complete, yet -- I hope -- also understandable. I tried to keep the notation simple. I moved the two other versions of the algorithm to the end of the document if you want to reinstall them.}

\subsection{Adaptive correction of the approximate posteriors}

\begin{figure}[htbp]
    \centering
    \includegraphics[width=0.8\linewidth]{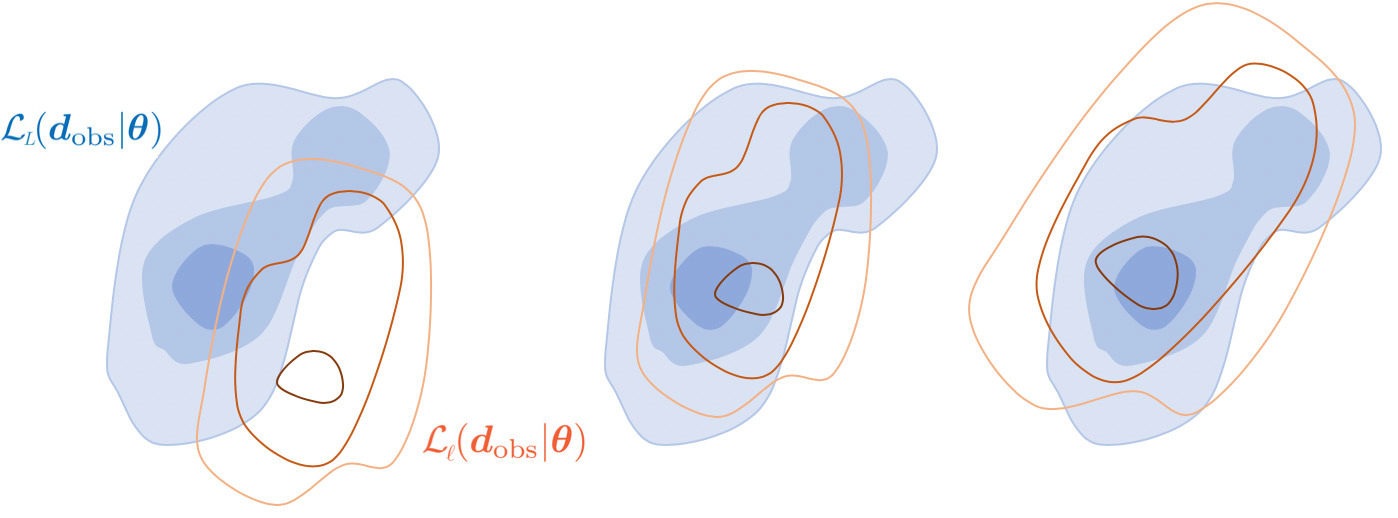}
    \caption{Effect of applying the Gaussian Adaptive Error Model (AEM). The first panel shows the initial state before adaptation, where the coarse likelihoods \rs{$\mathcal{L}_{\ell}({d}|\theta)$} (red/orange isolines) approximate the fine likelihood \rs{$\mathcal{L}_{L}({ d}|\theta)$} (blue contours) poorly. The second panel shows the effect of shifting the likelihoods by the mean of the bias. The third panel shows the effect of \rs{additionaly incorporating} estimates of the covariance of the bias. (Adapted from \cite{lykkegaard_accelerating_2020}.)}
    \label{fig:inflation_adaption}
\end{figure}

While the approach outlined above does guarantee sampling from the exact posterior, there are situations when convergence can be prohibitively slow. When the model approximation is poor, the \rs{delayed} acceptance probability is low, and many proposals are rejected. This will result in suboptimal acceptance rates and low effective sample sizes. The leftmost panel in Fig.~\ref{fig:inflation_adaption} shows a contrived example, where the approximate likelihoods (red/orange isolines) are offset from the likelihood on the finest level (blue contours) and their scales, shapes and orientations are incorrect.
Thus, as an additional modification, an Adaptive Error Model (AEM) is introduced to account for discrepancies between model levels. 

Let $\mathcal{F}_{\ell}$ denote a coarse forward map of level $\ell$ and $\mathcal{F}_{L}$ denote the forward map on the finest level~$L$. \rs{To obtain a better approximation of the data $d$ using $\mathcal{F}_{\ell}$, the two-level AEM suggested \rs{in \cite{kaipio_statistical_2007,cui_adaptive_2011} and analysed in \cite{cui_adaptive_2019}} is extended by adding a telescopic sum of the differences in the forward model output across all levels from $\ell$ to $L$}:
\begin{equation} \label{eq:inverse_problem_bias}
d = \mathcal F_L(\theta) + \epsilon = \mathcal F_{\ell}(\theta) + \mathcal B_{\ell}(\theta) + \epsilon \quad \mbox{with} \quad
\mathcal B_{\ell}(\theta) := \sum_{k=\ell}^{L-1} \underbrace{\mathcal F_{k+1}(\theta) -  \mathcal F_{k}(\theta)}_{:=B_k(\theta)}\,,
\end{equation}
\rs{denoting} the bias on level $\ell$ at ${\mathbf \theta}$. The trick in the context of MLDA is that, since $\mathcal{B}_{\ell}$ is just a simple sum, the individual bias terms $B_k$ from pairs of adjacent model levels can be estimated independently, so that new information can be exploited each time \textit{any} set of adjacent levels are evaluated for the same parameter value $\theta$.
Approximating each individual bias term $B_k = \mathcal F_{k+1} -  \mathcal F_{k}$ with a multivariate Gaussian $B^*_k \sim \mathcal N(\mu_k, \Sigma_k)$, \rs{the total bias $\mathcal{B}_\ell$ can be approximated by the Gaussian $\mathcal B^*_{\ell} \sim \mathcal N(\mu_{\mathcal{B}, \ell}, \Sigma_{\mathcal{B}, \ell})$ with $\mu_{\mathcal{B}, \ell} = 
\sum_{k} \mu_k$ and $\Sigma_{\mathcal{B}, \ell} = \sum_{k} \Sigma_k$.}

The \rs{bias-corrected} likelihood function for level $\ell$ is then proportional to
\begin{equation} \label{eq:likelihood_adaptive}
\mathcal{L}^*_{\ell}({d} | \theta) \propto \exp \left( -\frac{1}{2} 
\big(d -  \mathcal{F}_{\ell}(\theta) - \mu_\epsilon - \mu_{\mathcal{B}, \ell}\big)^T \big({\Sigma}_\epsilon + \Sigma_{\mathcal{B}, \ell} \big)^{-1} \big(d - \mathcal{F}_{\ell}(\theta) - \mu_\epsilon - {\mu}_{\mathcal{B}, \ell} \big) \right).
\end{equation}

\rs{One way to construct the AEM is} offline, by sampling from the prior before running the MCMC, as suggested in \cite{kaipio_statistical_2007}.
However, this approach requires a significant overhead prior to sampling, and may result in a suboptimal error model, since the bias in the posterior may \rs{differ substantially} from the bias in the prior. Instead, as suggested by \cite{cui_adaptive_2011}, an estimate for \rs{the 
$B_k$ can be constructed iteratively} during sampling, using the following recursive formulae 
for sample mean and sample covariance~\cite{haario_adaptive_2001}:
\begin{equation}
   {\mu}_{k,i+1}    = \frac{1}{i+1} \Big( i {\mu}_{k,i} + B_k(\theta^{i+1}) \Big) \quad \mbox{and}
\end{equation}
\begin{equation}
    {\Sigma}_{k,i+1} = \frac{i-1}{i} {\Sigma}_{k,i} + \frac{1}{i} \Big( i{\mu}_{k,i}\: { \mu}_{k,i}^T - (i+1) {\mu}_{k,i+1}\: { \mu}_{k,i+1}^T + B_k(\theta^{i+1}) \: B_k(\theta^{i+1})^T \Big)
\end{equation}
While this approach in theory compromises ergodicity in the strict sense, the recursively constructed sample moments exhibit \textit{diminishing adaptation} \cite{haario_adaptive_2001}.

\section{Implementation and Demonstration}\label{sec:pymc3}

The Multilevel Delayed Acceptance MCMC algorithm (Alg.~2) has been implemented in \texttt{PyMC3} \cite{salvatier_2016}, an open-source probabilistic programming package for Python built on top of the \texttt{Theano} library \cite{theano}. The code is available in the development version of \texttt{PyMC3}.\footnote{\href{https://github.com/pymc-devs/pymc3}{https://github.com/pymc-devs/pymc3}}. In the following section, we present a numerical experiment, in which we compare the ``vanilla'' MLDA sampler to the AEM-activated MLDA sampler. To demonstrate the effect of the AEM, we have chosen models of very low resolution on the coarse levels. It is important to stress, however, that the AEM is not a strict requirement for MLDA in cases, where the coarse models are better approximations of the fine.

\subsection{Example: Estimation of Soil Permeability in Subsurface Flow}
\label{sec:rocks}

In this example, a simple model problem arising in subsurface flow modelling is considered. Probabilistic uncertainty quantification is of interest in various situations, for example in risk assessment of radioactive waste repositories. Moreover, this simple PDE model is often used as a benchmark for MCMC algorithms in the applied mathematics literature. The classical equations which govern steady-state single-phase subsurface flow are Darcy's law coupled with an incompressibility constraint
\begin{equation}\label{eqn:fullDarcyEquations}
w + k\nabla p = g \quad \mbox{and} \quad \nabla \cdot w = 0, \quad \mbox{in} \quad D \subset \mathbb{R}^d
\end{equation}
for $d = 1,2$ or $3$, subject to suitable boundary conditions. Here $p$ denotes the hydraulic head of the fluid, $k$ the permeability tensor, $w$ the flux and $g$ is the source term.

A typical approach to treat the inherent uncertainty in this problem is to model the permeability as a random field $k = k(x,\omega)$ on $D \times \Omega$, for some probability space $(\Omega, \mathcal A, \mathbb P)$. Therefore, (\ref{eqn:fullDarcyEquations}) can be written as the following \rs{PDE with random coefficients}:
\begin{equation}\label{eq:spde}
-\nabla \cdot k(x,\omega)\nabla p(x,\omega) = f(x), \quad \mbox{for all} \quad x \in D,
\end{equation}
where $f:=-\nabla \cdot g$.
As a synthetic example, consider the domain $D := [0,1]^2$ with $f\equiv 0$ and deterministic boundary conditions
\begin{equation}
p|_{x_1=0} = 0, \quad p\vert_{x_1=1} = 1 \quad \mbox{and} \quad \partial_n p \vert_{x_2=0} = \partial_n p\vert_{x_2=1} = 0.
\end{equation}
A widely used model for the prior distribution of the permeability in hydrology is a log-Gaussian random field, characterised by the mean of $\log k$, here
chosen to be $0$, and by its covariance function, here chosen to be\vspace{-1ex}
\begin{equation}\label{eqn:covariance}
C({x},{y}) := \sigma^2 \exp\left(-\frac{\|{x} - {y}\|^2_2}{2\lambda^2}\right), \quad \mbox{for} \quad {x}, {y} \in D,
\end{equation}
with $\sigma = 2$ and $\lambda = 0.3$. The log-Gaussian random 
field is parametrised using a truncated Karhunen-Lo\`eve (KL) expansion 
of $\log k$, i.e., an expansion in terms of a finite set of independent, standard Gaussian random variables $\theta_i \sim \mathcal{N}(0,1)$, $i=1,\ldots,R$, given by
\begin{equation}
\log k(x,\omega) = \sum_{i=1}^R \sqrt{\mu_i} \phi_i({x})\theta_i(\omega).
\end{equation}
Here, $\{\mu_i\}_{i \in \mathbb N}$ are the sequence of strictly decreasing real, positive eigenvalues, and $\{\phi_i\}_{i\in \mathbb N}$ the corresponding $L^2$-orthonormal eigenfunctions of the covariance operator with kernel \rs{$C(x,y)$}. %(\ref{eqn:covariance}). 
Thus, the prior distribution on the parameter
$\theta = (\theta_i)_{i=1}^R$ in the stochastic PDE problem (\ref{eq:spde}) is $\mathcal{N}(0,I_R)$.

The aim is to infer the posterior distribution of $\theta$, conditioned on  measurements of $p$ at $M=25$ discrete locations $x^j \in D$, $j=1,\ldots,M$, stored in the vector ${d}_{obs} \in \mathbb R^{M}$. Thus, the forward operator is $\mathcal{F}:\mathbb{R}^R \to \mathbb{R}^M$ with $\mathcal{F}_j(\theta_\omega) = p(x^j,\omega)$.

\begin{figure}[htbp]
\centering
\includegraphics[width = 0.4\linewidth]{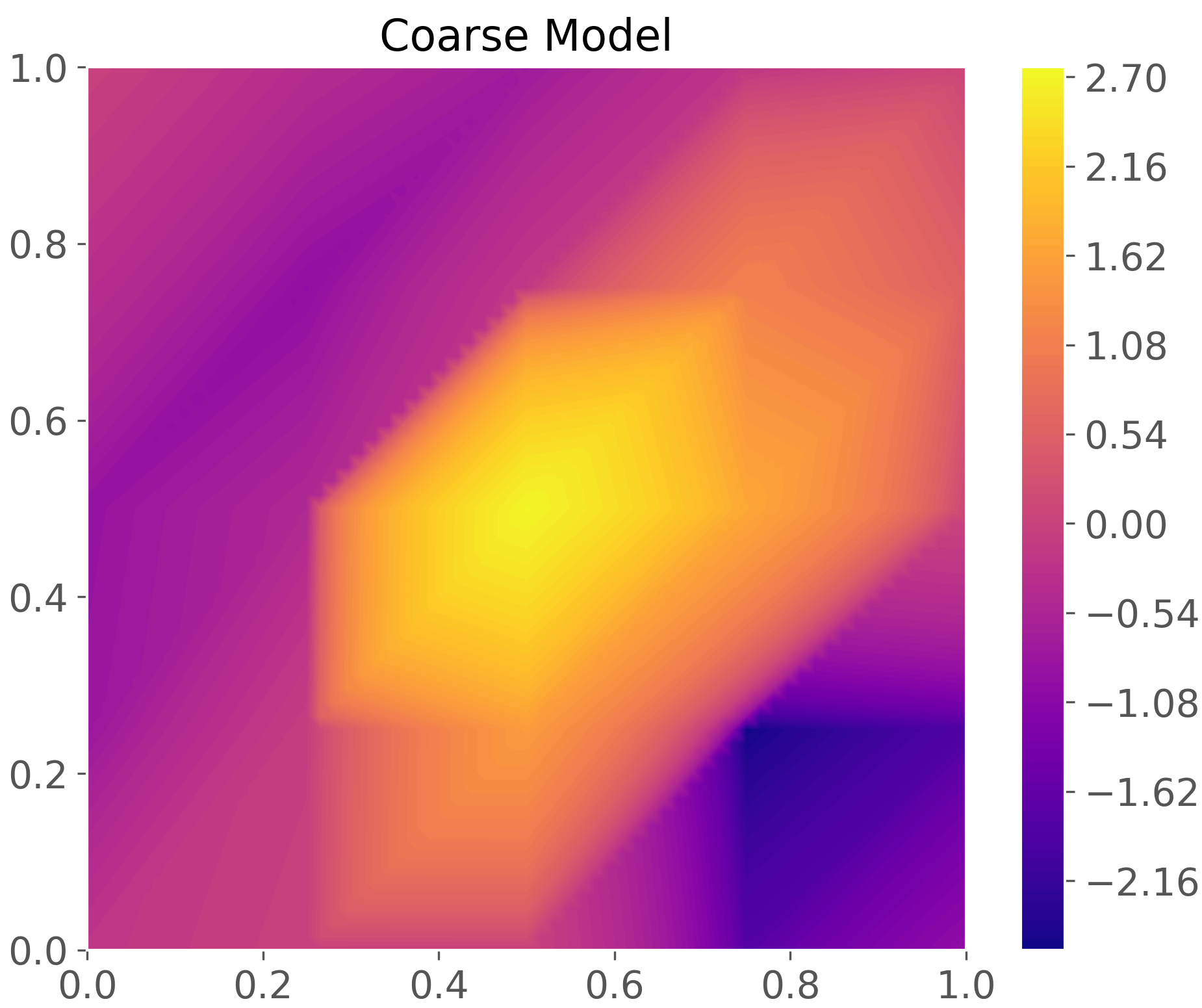}
\hspace{1cm}
\includegraphics[width = 0.4\linewidth]{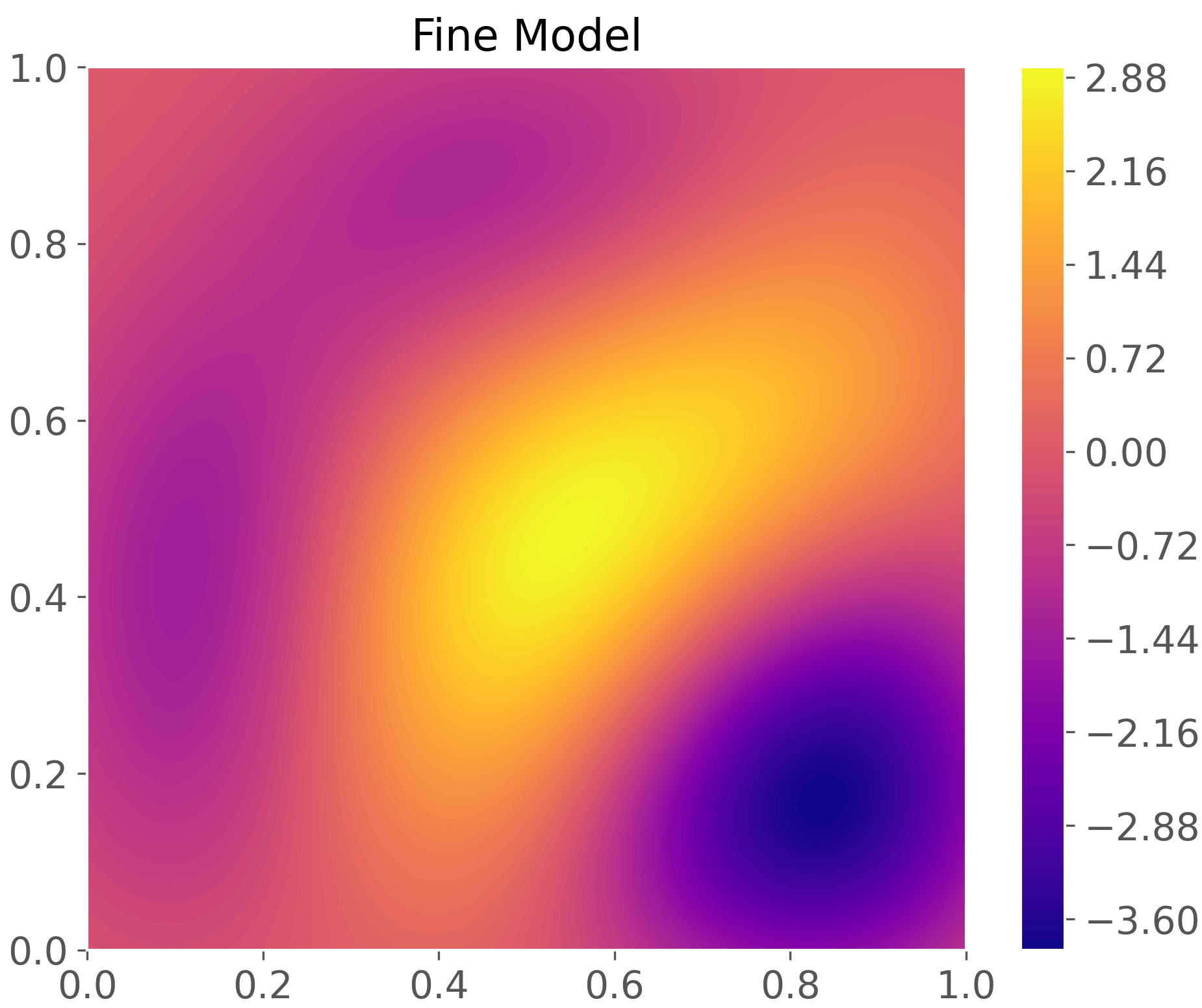}
\caption{True log-conductivity field of the coarsest model with $m_0$ grid points (left) and the finest model with $m_2$ grid points (right).}
\label{fig:gw_model}
\end{figure}

All finite element (FE) calculations were carried out with \texttt{FEniCS} \cite{langtangen_solving_2017}, using piecewise linear FEs on a uniform triangular mesh. The coarsest mesh $\mathcal T_0$ consisted of $m_0 = 5$ grid points in each direction, while subsequent levels were constructed by \rs{two steps of} uniform refinement of $\mathcal T_0$, \rs{leading to} $m_\ell = 4^\ell(m_0 -1) + 1$ grid points in each direction \rs{on the three grids $\mathcal{T}_\ell$, $\ell = 0, 1, 2$ (Fig.~\ref{fig:gw_model})}. 

\rs{To demonstrate the excellent performance of MLDA with the AEM}, synthetic data was generated by drawing a sample $\theta^{ex}$ from the prior distribution and solving (\ref{eq:spde}) with the resulting realisation of $k$ on $\mathcal T_2$. To construct $d_{obs}$, the computed discrete hydraulic head values at $(x^j)_{j=1}^M$, were then perturbed by independent Gaussian random variables, i.e. by a sample \rs{$\epsilon^* \sim \mathcal N(0, \Sigma_\epsilon)$ with $\Sigma_\epsilon = 0.01^2 I_M$}.

To compare the ``vanilla'' MLDA approach to the AEM-\rs{enhanced} version, we sampled the same model using identical sampling parameters, with and without AEM activated. For each approach, we sampled four independent chains, each initialised at a random point from the prior. For each independent chain, we drew 5000 samples plus a burn-in of 2000. We used \rs{subchain lengths $J_0 = J_1 =5$}, since that produced the best trade-off between computation time and effective sample size \rs{for MLDA with the AEM. Note that the cost of computing the subchains on the coarser levels only leads to about a 50\% increase in the total cost for drawing a sample on level $L$. The \texttt{PyMC3} non-blocked Random Walk Metropolis Hastings (RWMH) sampler was employed on the coarsest level with automatic step-size tuning during burn-in to achieve an acceptance rate between 0.2 and 0.5. All other sampling parameters were maintained at the default setting of the \texttt{MLDA} method.}

\begin{figure}[htbp]
    \centering
    \includegraphics[width = 0.9\linewidth]{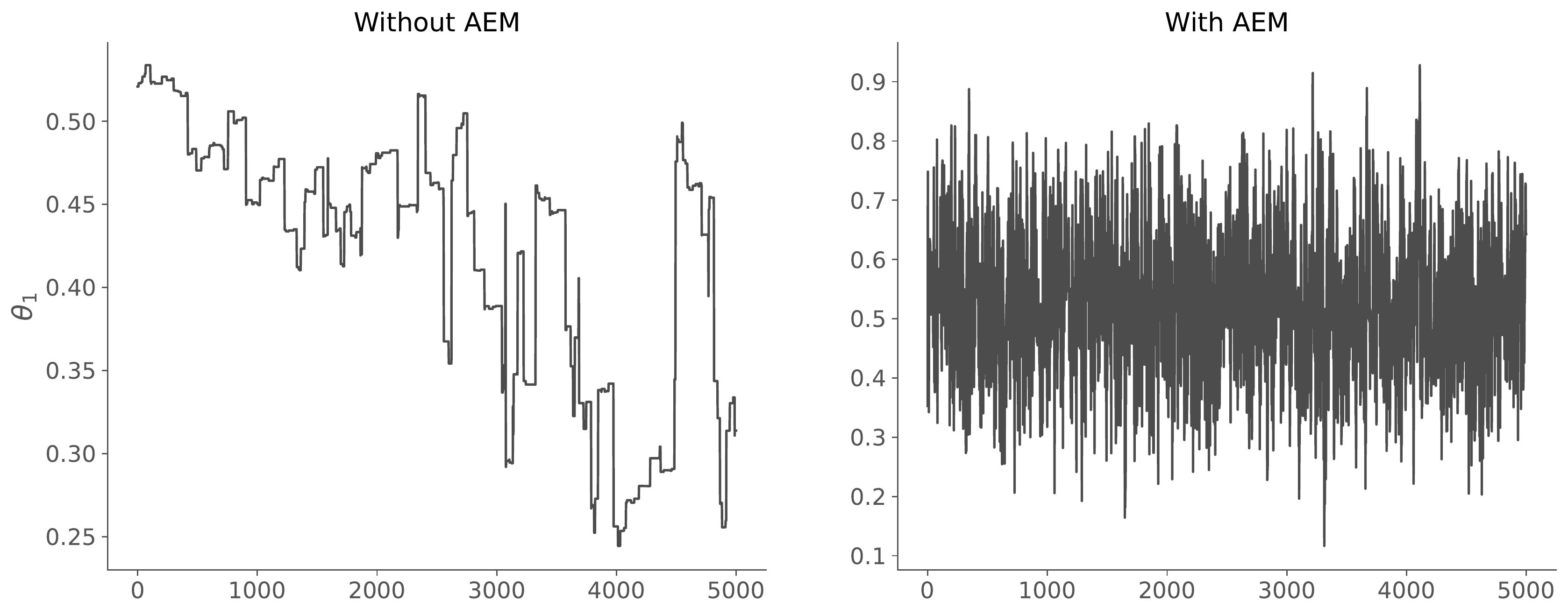}
    \caption{Traces of $\theta_1$ on level $\ell = 2$, for MLDA without (left) and with AEM (right).}
    \label{fig:gw_traces}
\end{figure}

\rs{To assess the performance of the two approaches the Effective Sample Size (ESS) for each parameter was computed  \cite{vehtari_rank-normalization_2020}.} Since the coarsest model was quite a poor approximation of the finest, running \rs{MLDA} without the Adaptive Error Model (AEM), yielded very poor results. None of the four chains converged, there was poor mixing, a sub optimal acceptance rate of 0.019 on level $L$, and an ESS of 4 out of 20000 samples, meaning that each independent chain was only capable of producing a single independent sample.
When the AEM was employed and otherwise using the exact same sampling parameters, we observed convergence for every chain, good mixing, an acceptance rate of 0.66 on level $L$ and an ESS of 3319 out of 20000 samples (Fig. \ref{fig:gw_traces}). In comparison, a single-level non-blocked RWMH sampler on grid $\mathcal{T}_2$ with automatic step-size tuning during burn-in produced an ESS of 19 out of 5000 samples with an acceptance rate of 0.26.

\rs{Note that the particular numerical experiment was chosen to demonstrate the dramatic effect that employing the AEM can have in MLDA. Thus, making it possible to use multilevel sampling strategies with very crude approximate models. A FE mesh with 25 degrees of freedom is extremely coarse for a Gaussian random field with correlation length $\lambda=0.3$, yet using the AEM it still provides an excellent surrogate for delayed acceptance. Typically much finer models are used in real applications with longer subchains on the coarser levels (cf.~\cite{dodwell_hierarchical_2015}). The AEM will be less critical in that case and MLDA will also produce good ESS without the AEM.
%\ml{As mentioned in the introduction to this section, the demand for employing the AEM arises from the low-resolution coarse models. However, given better coarse model approximations, MLDA is capable of producing optimal sampling rates without AEM. 
In a future journal paper, this topic will be carefully studied along with a comparison with other samplers on the finest level and an analysis of the multilevel variance reduction capabilities of MLDA.}

\section*{Broader Impact}
This research has the potential to make unbiased uncertainty quantification of expensive models available to a greater audience, including engineers employed in risk assessment and reliability engineering. Since many engineering problems involve solving PDEs, multi-level hierarchies can easily be introduced using grid refinement, making this method exceptionally well suited for engineering applications.

\section*{Acknowledgements}
The work was funded by a Turing AI fellowship (2TAFFP\textbackslash100007) and the Water Informatics Science and Engineering Centre for Doctoral Training (WISE CDT) under a grant from the Engineering and Physical Sciences Research Council (EPSRC), grant number EP/L016214/1. 

\bibliographystyle{unsrt}  
\bibliography{main}

\end{document}